\documentclass[trackchanges,twocolumn]{aastex701}

\begin{document}


\title{Discovery of an Accretion Burst in a Free-Floating Planetary-Mass Object}

\correspondingauthor{Victor Almendros-Abad}
\email{victor.almendrosabad@inaf.it}

\author[0000-0002-4945-9483]{V. Almendros-Abad}
\affiliation{Istituto Nazionale di Astrofisica (INAF) - Osservatorio Astronomico di Palermo, Piazza del Parlamento 1, 90134, Palermo, Italy}
\email{victor.almendrosabad@inaf.it}

\author[0000-0001-8993-5053]{Aleks Scholz}
\affiliation{SUPA, School of Physics \& Astronomy, University of St Andrews, North Haugh, St Andrews, KY169SS, United Kingdom}
\email{as110@st-andrews.ac.uk}

\author[0000-0002-2234-4678]{Belinda Damian}
\affiliation{SUPA, School of Physics \& Astronomy, University of St Andrews, North Haugh, St Andrews, KY169SS, United Kingdom}
\email{bd64@st-andrews.ac.uk}

\author[0000-0001-5349-6853]{Ray Jayawardhana}
\affiliation{Department of Physics \& Astronomy, Johns Hopkins University,  Baltimore, MD, 21218, USA}
\email{rayjay@jhu.edu}

\author[0000-0001-7868-7031]{Amelia Bayo}
\affiliation{European Southern Observatory, Karl-Schwarzschild-Strasse 2, D-85748 Garching bei München, Germany}
\email{ameliamaria.bayoaran@eso.org}

\author[0000-0001-6362-0571]{Laura Flagg}
\affiliation{Department of Physics \& Astronomy, Johns Hopkins University,  Baltimore, MD, 21218, USA}
\email{laura.s.flagg@gmail.com}

\author[0000-0002-7989-2595]{Koraljka Mu\v{z}i\'c}
\affiliation{Instituto de Astrof\'{i}sica e Ci\^{e}ncias do Espaço, Faculdade de Ci\^{e}ncias, Universidade de Lisboa, Ed. C8, Campo Grande, 1749-016 Lisbon, Portugal}
\email{koralja@gmail.com}

\author[0000-0002-4608-7995]{Antonella Natta}
\affiliation{School of Cosmic Physics, Dublin Institute for Advanced Studies, 31 Fitzwilliam Place, Dublin 2, Ireland}
\affiliation{University College Dublin (UCD), Department of Physics, Belfield, Dublin 4, Ireland}
\email{antonella.natta@gmail.com}

\author[0000-0001-8764-1780]{Paola Pinilla}
\affiliation{Mullard Space Science Laboratory, University College London, Holmbury St Mary, Dorking, London, UK}
\email{p.pinilla@ucl.ac.uk}

\author[0000-0003-1859-3070]{Leonardo Testi}
\affiliation{Dipartimento di Fisica e Astronomia, Università di Bologna, Via Gobetti 93/2, 40122, Bologna, Italy}
\email{leonardo.testi@unibo.it}

\begin{abstract}
We report the discovery of a long-lasting burst of disk accretion in Cha J11070768-7626326 (Cha 1107-7626), a young, isolated, 5-10 M$_{\mathrm{Jupiter}}$ object. In spectra taken with XSHOOTER at ESO's Very Large Telescope as well as NIRSPEC and MIRI on the James Webb Space Telescope, the object transitions from quiescence in April-May 2025 to a strongly enhanced accretion phase in June-August 2025. The line flux changes correspond to a 6–8-fold increase in the mass accretion rate, reaching $10^{-7}$\,M$_{\mathrm{Jupiter}}$yr$^{-1}$, the highest measured in a planetary-mass object. During the burst, the H$\alpha$ line develops a double-peaked profile with red-shifted absorption, as observed in stars and brown dwarfs undergoing magnetospheric accretion. The optical continuum increases by a factor of 3-6; the object is $\sim$1.5-2\,mag brighter in the R-band during the burst. Mid-infrared continuum fluxes rise by 10-20\%, with clear changes in the hydrocarbon emission lines from the disk. We detect water vapour emission at 6.5-7$\,\mu m$, which were absent in quiescence. By the end of our observing campaign, the burst was still ongoing, implying a duration of at least two months. A 2016 spectrum also shows high accretion levels, suggesting that this object may undergo recurring bursts. The observed event is inconsistent with typical variability in accreting young stars and instead matches the duration, amplitude and line spectrum of an EXor-type burst, making Cha1107-7626 the first substellar object with evidence of a potentially recurring EXor burst.
\end{abstract}



\section{Introduction} 
\label{sec:intro}

Deep surveys in star forming regions have identified young free-floating objects with masses below the Deuterium burning limit at 13 Jupiter masses, or 0.013 $M_\odot$, starting with the pioneering studies by \citet{zapatero2000} and \citet{lucas2000}. In terms of their masses, these objects are comparable to giant planets. They share some features with planets, including atmospheric properties  \citep{bonnefoy2014}. In contrast to planets, they are found in isolation, not in orbit around a star, and harbor accreting disks at young ages \citep{luhman2005, jaya2006} The existence of free-floating planetary-mass objects (FFPMOs) raises many fundamental questions. Are these the lowest-mass objects formed like stars? What is the low-mass limit for star formation? Or are these giant planets that have been ejected from their planetary systems? How does the ejection occur exactly? \citep{scholz2022,miret2023,langeveld2024}

Observationally, the most obvious path to constraining the nature of FFPMOs is a detailed characterization of the objects and their environment. Here, the study of disks and accretion is particularly relevant. The presence of a disk and ongoing accretion is typically interpreted as a sign that the object shares a formation and early evolutionary path with stars. Disks have been found through infrared excess for objects with 5-10 Jupiter masses \citep{testi2002,luhman2008}. The disk fractions do not seem to decline in the planetary-mass domain \citep{seo2025}. In some cases, evidence for gas accretion has been seen \citep{viswanath2024}. 

The object Cha1170-7626 is a key target in this regard - with an estimated mass of only 5-10 Jupiter masses, it is one of the lowest-mass FFPMO with clear evidence for disk and accretion \citep{luhman2008}. Recent observations with VLT and JWST have shown a) clear evidence for infrared excess from 4 to 12 $\mu$m, b) a silicate emission feature at 10 $\mu$m similar to those seen in stars and brown dwarfs, c) hydrocarbon emission lines, indicating a carbon-rich disk chemistry, and d) multiple accretion-induced emission lines  \citep{flagg2025,damian2025}. These characteristics make Cha1107-7626 the posterchild for disk accretion in the planetary-mass domain. 

Using observations from VLT/XSHOOTER and instruments on-board JWST we recently discovered an accretion burst in Cha1107-7626, which began in June 2025 and was still ongoing by the end of August 2025. This is the first time an accretion burst is seen in an object with such a low mass. In Section \ref{sec:obs} we present the observations and the resulting spectra. In Section \ref{sec:line} we analyse the changes in the accretion lines during the burst, derive the corresponding changes in mass accretion rate, and discuss the change in line profiles. In Section \ref{sec:sed} we examine any changes in the broad spectral energy distribution in the optical, near- and mid-infrared. Our results are summarized in Section \ref{sec:conc}.

\begin{figure*}
    \centering
    \includegraphics[width=0.95\textwidth]{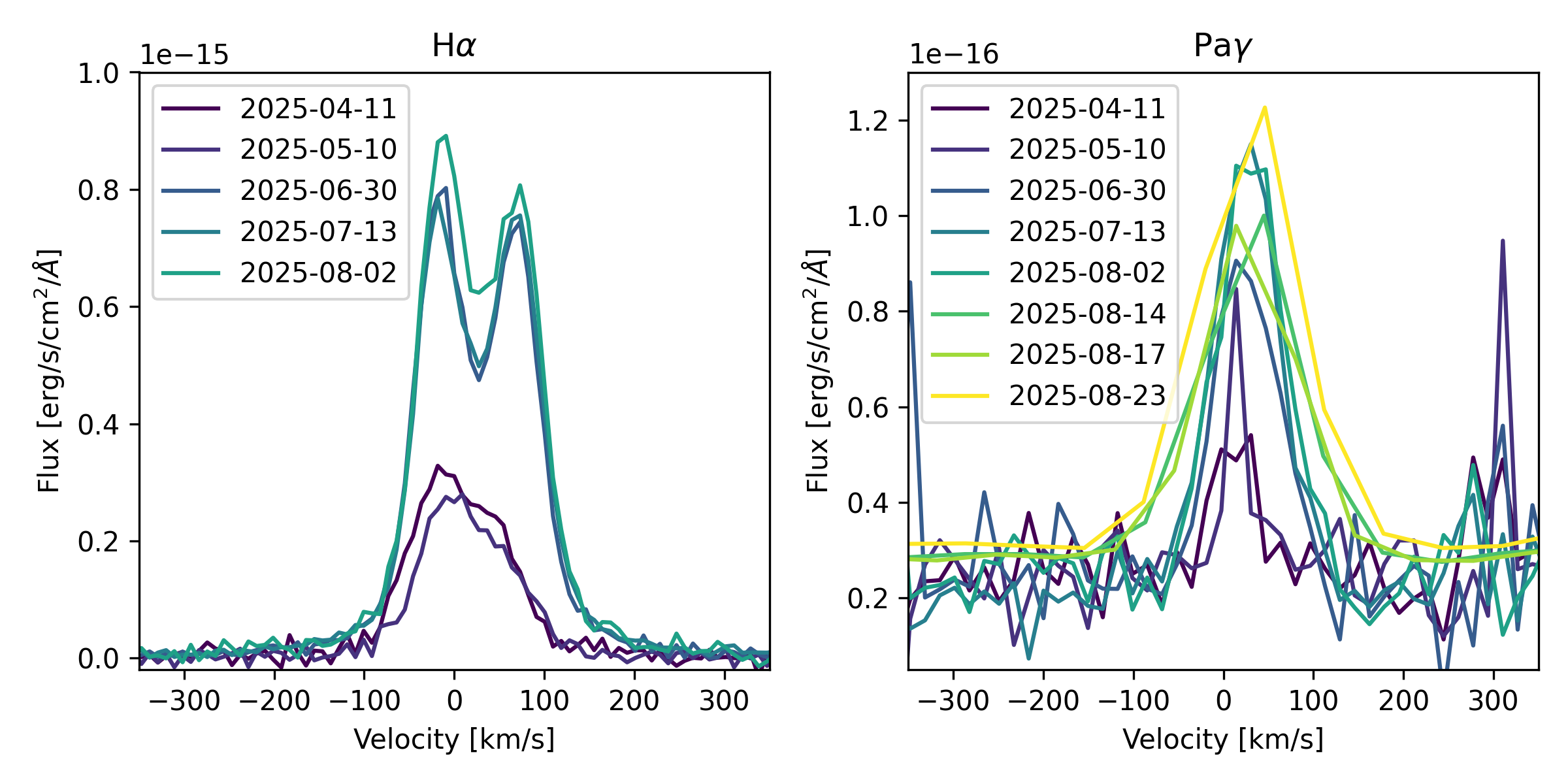}
    \caption{Line profile changes in H$\alpha$ from VLT/XSHOOTER (left) and Paschen $\gamma$ from both XSHOOTER and JWST (right) seen in VLT/XSHOOTER observations of Cha1107-7626. The eight epochs are colour-coded and clearly show the evolution of the burst.} 
    \label{fig:lines}
\end{figure*}

\section{Observations} 
\label{sec:obs}

\subsection{X-SHOOTER}

The burst was discovered as part of observations conducted with the XSHOOTER spectrograph on ESO's VLT (program 115.2850). Initially three epochs were obtained, in April, May, and late June 2025, the last one showing a clear increase in the strength of accretion-related emission lines. Two further epochs were obtained in mid July and early August in Director's Discretionary Time (programs 115.29FC and 115.29G3). All spectra were obtained under clear conditions except for the first, which ended with thick cloud conditions but yielded quality comparable to the second epoch. Seeing was $\leq$1.1$''$ in all epochs except late June 2025, when it was slightly worse.

XSHOOTER is a medium-resolution spectrograph that offers a very broad wavelength coverage from the ultraviolet to the near infrared. All five observations used the same setup, with slit widths of 1.0$''$/0.9$''$/0.9$''$ in the UVB/VIS/NIR arms. Narrow-slit exposures were 600 s (UVB), 900 s (VIS), and 7×150 s (NIR), accompanied by 5$''$ wide-slit exposures (10\% of narrow-slit time) for slit-loss correction. In July and August, high airmass ($>2.0$) limited the observing window and prevented wide-slit observations. Data reduction was performed with the ESO XSHOOTER pipeline v.3.6.8 \citep{modigliani2010} with default settings, and telluric correction using Molecfit \citep{smette2015} v.4.4.2. Flux calibration was carried out by comparing the narrow-slit spectra to the corresponding wide-slit spectra in the first three epochs \citep{alcala2017}. The wide-slit fluxes were found to be consistent across these epochs within 10\%. Since the observing conditions were similar for the June to August 2025 epochs, we used the wide-slit spectrum taken in June to also calibrate the July and August 2025 observations.

\subsection{JWST}

We observed Cha1107-7626 three times with JWST between the 14 and 22 of August 2025, using 9.7\,h of Director's Discretionary Time (program 9448). For each epoch we used NIRSPEC \citep{jakobson2022} to obtain a low-resolution spectrum from 0.6 to 5.0$\,\mu$m (with the PRISM setup) as well as medium-resolution spectra for two windows at 1.0 to 1.8$\,\mu$m and 1.7 to 3.0$\,\mu$m (using grisms G140H and G235H). In the first epoch, we also obtained a low-resolution spectrum with MIRI/LRS \citep{wright2023} from 5 to 12$\,\mu$m. The on-source times were 525\,sec for PRISM, 1838\,s for the mid-resolution NIRSPEC data, and 3591\,s for MIRI-LRS. In the following, we used the pipeline reduced data which shows excellent consistency between setups. We also make use of the archival data for the same target obtained in August 2024 in program 4583 \citep{flagg2025,damian2025}. For the new PRISM and MIRI-LRS observations, we closely followed the configuration that was successfully used in the program from 2024. 

\section{Line emission}
\label{sec:line}

\subsection{Evolution of the line spectrum}

The line spectrum for Cha1107-7626 shows a clear transition from a quiescent state to strongly enhanced accretion, which is sustained over several months. This is the hallmark of a long-lasting accretion burst \citep{fischer2023}.

In Figure \ref{fig:lines} we illustrate the evolution of two accretion-related emission lines through this event: Balmer H$\alpha$ and Paschen $\gamma$. In April and May 2025, Cha1107-7626 exhibits relatively weak accretion-related lines in the Hydrogen Balmer and Paschen series. 
Between June and August, the emission line spectrum clearly strengthens. This is evident in the H$\alpha$ line which becomes broader, stronger, and also develops a double-peak profile, as seen in Figure \ref{fig:lines}. Such an enhancement in the emission line strength is also apparent when comparing NIRSPEC/PRISM spectra from August 2024 and August 2025, despite the very low resolution -- the more recent observations exhibit a much stronger H$\alpha$ (see Section \ref{sec:sed_cont}).

\begin{figure*}
    \centering
    \includegraphics[width=0.95\textwidth]{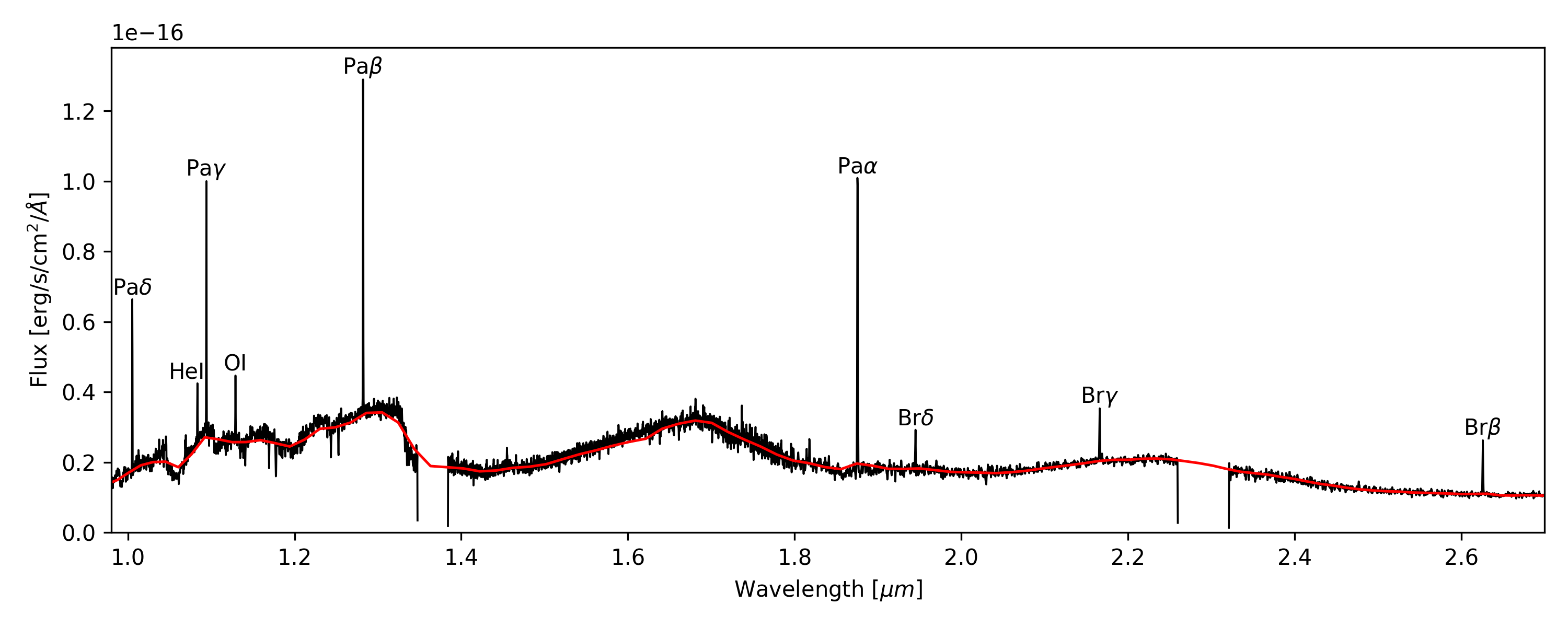}
    \caption{Medium-resolution G140H and G235H JWST NIRSPEC spectra from 14/08/2025 (black line) together with the PRISM spectrum from the same epoch (red line). The strongest emission lines detected are identified.} 
    \label{fig:lines_jwst}
\end{figure*}

Analogous changes are observed in other hydrogen lines. The Paschen $\gamma$ line significantly increases in width and flux (Figure \ref{fig:lines}). In addition, lines not seen in April and May are visible in the June-August epochs, including Brackett $\gamma$, Paschen $\beta$, the Calcium infrared triplet and higher Paschen lines. The higher-resolution JWST/NIRSPEC observations also reveal strong emission from Paschen $\alpha$, Brackett $\beta$, and several higher Brackett and Pfund lines (Figure \ref{fig:lines_jwst}). The Balmer $\beta$ line also strengthens in July and August 2025. From June to August 2025, the characteristics of the emission line spectrum show only minor changes.

\subsection{Double-peaked H$\alpha$}

The double-peaked, asymmetric H$\alpha$ profile seen June to August 2025 is in itself an exciting discovery. Such a profile is caused by absorption, superimposed onto the broad emission, shifted slightly to the red. In this particular case, the absorption trough is redshifted by 20-40 kms$^{-1}$. This type of line profile is a hallmark of funneled accretion seen at high inclination: The red-shifted absorption is caused by cool, infalling gas in the accretion column which is seen projected against the hot spot and causes the broad emission \citep{bouvier2003}. 

A very similar H$\alpha$ profile change was observed in 2005 for the accreting brown dwarf TWA27 \citep{scholz2005,scholz2006}. In that case, the redshifted absorption was modulated on a rotational timescale. For Cha1107-7626 we do not have the high cadence data to check for this modulation. Similar to Cha1107-7626, the appearance of the double-peaked profile in TWA27 was associated with a strong increase in the strength and width of the line. The profile changes in TWA27 were interpreted as clear evidence for magnetically channeled accretion in substellar objects. The spectra for Cha1107-7626 presented in this paper suggest that the same mode of accretion can be at work at objects with masses comparable to those of giant planets. 

More specifically, we may observe a switch between two different accretion modes, for example, the stable/unstable modes suggested by \citet{kurosawa2013}. Stable accretion means that the configuration of the accretion columns is consistent over many rotational cycles. On the other hand, in unstable mode the accretion columns are variable on short timescales. This leads to enhanced accretion rates, as well as persistent redshifted absorption in lines -- as observed in our target. 

\subsection{Accretion rate changes}

The accretion luminosity at each epoch was estimated following the same method to \citet{flagg2025}, using established recipes. We adopted the photospheric parameters derived in \citet{damian2025}. In short, we measure the line luminosities on the de-reddened spectra, and convert to the total accretion luminosity using the empirical relations from \citet{alcala2017}.

\begin{figure*}
    \centering
    \includegraphics[width=0.7\textwidth]{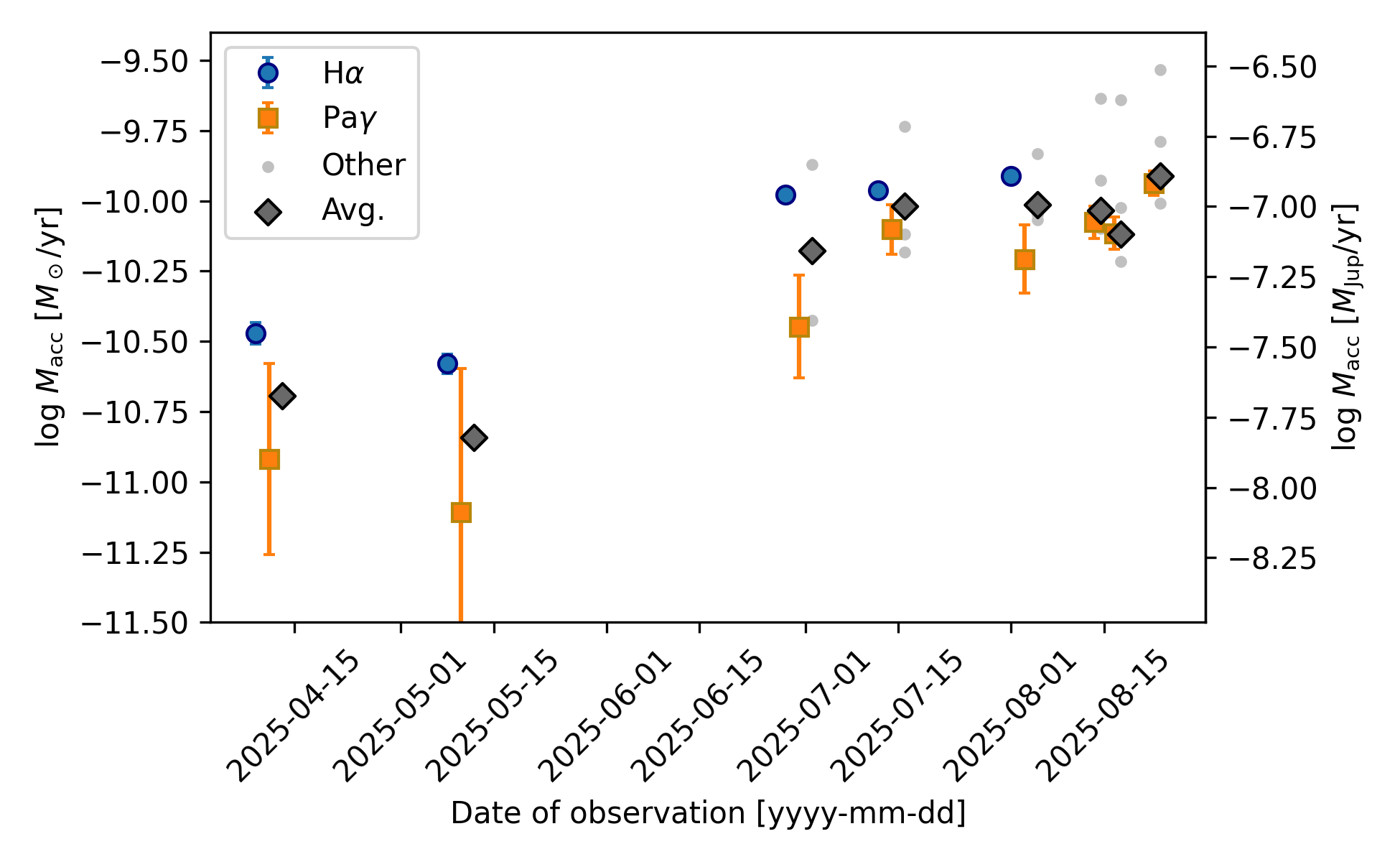}
    \caption{Time evolution of the mass accretion rate across all observations seen in H$\alpha$ (blue circles), Paschen $\gamma$ (orange squares). Black diamonds show the average over the following available tracers at each epoch: Paschen $\delta$, Paschen $\beta$, Brackett $\gamma$, and H$\alpha$, each shown with small light gray circles. A small offset in date has been applied for clarity. The Pa$\gamma$ and H$\alpha$ accretion rate error bars represent the uncertainties from the line flux measurements only. Systematic uncertainties in the accretion rate, primarily from the adopted $L_{\rm line}$–$L_{\rm acc}$ relation and stellar parameters, are not shown; these amount to $\sim$0.4 dex.} 
    \label{fig:accretion}
\end{figure*}

The evolution of the mass accretion rate from H$\alpha$ and Paschen $\gamma$ (i.e. the two lines detected across all XSHOOTER epochs) is shown in Figure \ref{fig:accretion}. We also show the average accretion rate from several lines detected in the XSHOOTER and JWST observations: H$\alpha$, Paschen $\delta$, Paschen $\gamma$, Paschen $\beta$ and Brackett $\gamma$. We note that H$\alpha$ is not covered in the JWST grism spectra used for the accretion rate estimates, while Paschen $\delta$, Paschen $\beta$, and Brackett $\gamma$ are detected only during the enhanced accretion phase. Both the individual line tracers and the average indicate an 6-fold increase (corresponding to $\sim$0.8 dex) between the quiescence and enhanced phases. We estimate that self-absorption in H$\alpha$ during the elevated state lowers the measured accretion rate by 10\%, implying a true increase closer to 8-fold. While the errors in the accretion rates derived from individual tracers are substantial, they are primarily systematic, which means that our estimate of the {\it change} in mass accretion rate over time is robust, as further supported by the consistency among multiple lines. In the last JWST epoch we measure the strongest accretion rate, showing that the burst is still ongoing. This implies a duration of the burst of more than two months.

The accretion rate during the burst is $\sim10^{-10}$\,M$_{\odot}$yr$^{-1}$, or $10^{-7}$\,M$_{\mathrm{Jup}}$yr$^{-1}$. This represents the strongest accretion rate measured on a planetary-mass object \citep[see][]{betti2023}. The only comparable measurement to our knowledge is that of OTS 44 derived from Paschen $\beta$ \citep{joergens2013}, which contrasts with the H$\alpha$ measurement reported in the same paper that is two orders of magnitude lower. Those diagnostics were not obtained simultaneously, so the discrepancy may well be explained by strong accretion variability, similar to what we observe in Cha1107-7626. Compared to measured accretion rates in embedded protoplanets, the accretion rate in Cha1107-7626 during the burst is higher by at least a factor of $\sim 20$ \citep{close2025}.

The first two XSHOOTER epochs are consistent with the lower accretion rates measured from H$\alpha$ \citep{flagg2025} in a 2008 optical spectrum \citep{luhman2008}, which may represent the quiescent state. On the other hand, the XSHOOTER and JWST grism epochs from June to August 2025 match the accretion rates measured from Paschen $\beta$ \citep{flagg2025} in a near-IR spectrum taken in 2016 \citep{almendros2022}. This suggests that events like the one reported here are recurring in this particular object, on timescales of years.

If we instead use line–accretion luminosity relations from ''accretion-shock'' models \citep{aoyama2021} we obtain a similar temporal evolution, but the accretion rates in all epochs and lines are higher by about one order of magnitude. Thus, when adopting this specific model, the accretion rate during the burst is $10^{-9}$\,M$_{\odot}$yr$^{-1}$, or $10^{-6}$\,M$_{\mathrm{Jup}}$yr$^{-1}$. 

\section{Spectral energy distribution}
\label{sec:sed}

\subsection{Continuum}
\label{sec:sed_cont}

\begin{figure*}
    \centering
    \includegraphics[width=\textwidth]{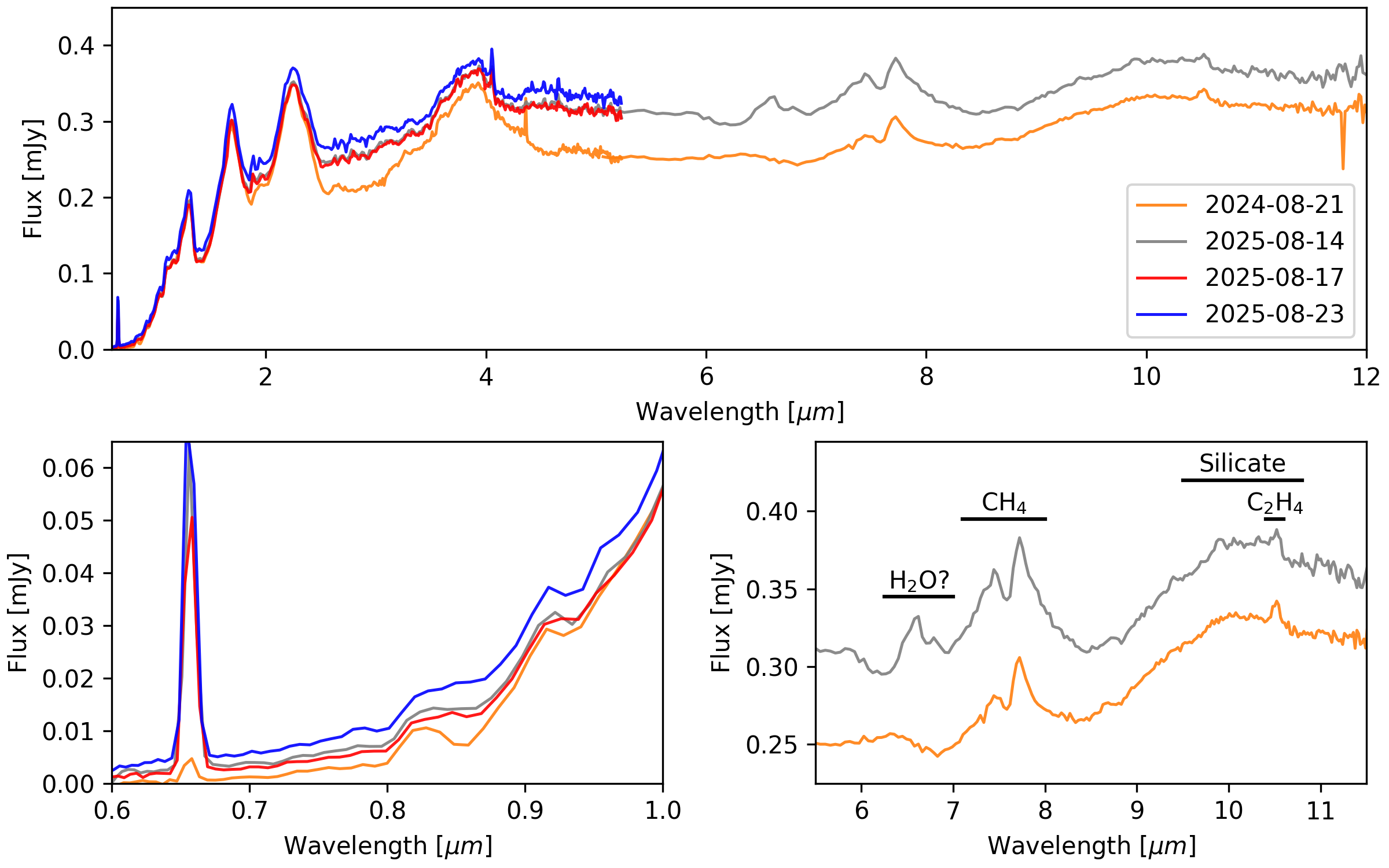}
    \caption{Time evolution of the spectral energy distribution for Cha1107-7626, as observed with JWST NIRSPEC-PRISM and MIRI-LRS, in August 2024 and August 2025. Bottom left and right panels show a zoom to the optical and mid-infrared, respectively. The main mid-infrared features discussed in Section \ref{sec:sed} are indicated in the bottom right panel.} 
    \label{fig:sed}
\end{figure*}

The JWST data from August 2024 and August 2025 provides us with an opportunity to probe the spectral energy distribution of Cha1107-7626 in quiescence and during the burst, over a wide wavelength range. In Figure \ref{fig:sed} we show all available epochs observed with NIRSPEC-PRISM and with MIRI-LRS. The object shows enhanced continuum emission during the accretion burst. In the optical and outside the H$\alpha$ line the flux levels are elevated by about a factor of 3-6 when comparing August 2025 to August 2024. Combined with the increase in H$\alpha$, this would imply a photometric amplitude of about 1.5-2\,mag in the R-band. The additional continuum emission in this wavelength range likely originates from the accretion shock, so-called 'veiling' \citep{calvet1998}. We note that such variations are not detected in the XSHOOTER spectra due to the low S/N in the continuum at these wavelengths.

In the near infrared range from 1-2$\,\mu$m, the continuum emission remains approximately constant across all NIRSPEC and XSHOOTER observations, with the exception of the last NIRSPEC epoch, which shows increased fluxes by about 10\%. At longer wavelengths, however, we see elevated flux levels by 10-20\% in all 2025 epochs, most strikingly between 5 and 10$\,\mu$m. The straightforward explanation for the enhanced mid-infrared fluxes is an increase in the temperature of the inner disk, caused by the additional heating from the additional accretion \citep{fischer2023}. Only a marginal temperature increase is needed for 20\% in flux increase -- assuming blackbody emission and typical inner disk temperatures of 500-1000\,K, the required temperature increase is less than 100\,K for the wavelength range in question.

The characteristic of the continuum changes are the expected consequence of an increase in the mass accretion rate, see the modeling in \citet{scholz2013} for an exploration of this issue. For a moderate change in the accretion rate, the near-infrared fluxes remain almost unchanged. The optical flux, however, is strongly affected by the additional emission from the accretion shock, while the mid-infrared responds to the additional heating in the disk.

\subsection{Hydrocarbon lines}

As reported by \citet{flagg2025}, Cha1107-7626 shows clear hydrocarbon emission lines originating in the disk in its quiescent mid-infrared spectrum. In particular, emission from CH$_4$ at 7-8$\,\mu$m and C$_2$H$_4$ at 10.5$\,\mu$m was identified based on the 2024 JWST data. The new MIRI data from August 2025 shows again both features, confirming that this is a carbon-rich disk. 

Interestingly, we observe clear changes in the molecular line spectrum compared to quiescence. The double-peaked feature at 7-8$\,\mu$m has changed in shape; in the new spectrum, the blue peak has increased in strength relative to the red one. In addition to these hydrocarbon lines, the burst spectrum shows a feature around 6.6$\,\mu$m, which was not visible in quiescence (bottom right panel Figure \ref{fig:sed}). The emission is just beyond the 6.3\,$\mu$m absorption caused by photospheric water \citep{cushing2006}. A similarly shaped 6.6$\,\mu$m feature has been reported in Spitzer spectra of accreting low-mass stars and was attributed to water vapour \citep{sargent2014}. More subtle excesses in the same wavelength region have also recently been associated with water vapour in JWST/MIRI spectra of very low-mass stars and a substellar object \citep{xie2023,arabhavi2025,morales2025}. The strongest rovibrational H$_2$O bands predicted for accreting young stars are located near this wavelength \citep{banzatti2025}, and increased H$_2$O emission has been observed during accretion bursts in EX Lupi \citep{smith2025}. While we note that C$_2$H$_6$ also has transitions in this spectral region \citep{henning2024}, the empirical evidence and the physical context of an accretion-driven temperature increase both favor an H$_2$O origin. We therefore identify the 6.6 $\mu$m feature as water vapour. This is the first time the chemical changes in the disk caused by increased accretion are observed in a substellar object.  

\subsection{Silicate feature}

Cha1107-7626 exhibits a clear silicate feature at 8-10$\,\mu$m, likely caused by a mix of small amorphous as well as crystallized grains \citep{damian2025}. By eye, the shape of the silicate feature in 2025 is very similar to 2024. We calculated the standard metrics to quantify the strength and shape of the silicate feature, after dereddening, as shown in \citet{damian2025}. In 2024, the peak-over-continuum ratio at 10$\,\mu$m was 1.2$\pm$0.05, while the continuum normalized flux ratio at between 11.3 and 9.8$\,\mu$m was 0.87$\pm$0.03. For comparison, the numbers for the August 2025 spectrum are 1.18$\pm$0.05 and 0.88$\pm$0.03. This shows that the strength and shape of the silicate feature before and during the burst is very similar.

For reference, strong changes in the silicate feature have been observed during accretion bursts for T Tauri stars. In particular, for EX Lupi the silicate feature showed notable changes when observed two months after the peak of the outburst in 2008, compared to quiescence \citep{abraham2009,juhasz2012}. While in quiescence the silicate profile looked almost triangular, typical for ISM-type grains, it became more complex and multi-peaked during the late stages of the outburst. The 2008 outburst of EX Lupi was an extreme event for this prototypical outbursting star, with a duration of 7 months, a brightness increase by 5\,mag, and accretion rate changes by a factor of 30 \citep{aspin2010}. It is therefore not directly comparable to the event reported here for Cha1107-7626.

\section{Summary and outlook}
\label{sec:conc}

In this paper we present clear evidence for a significant accretion burst in the free-floating planetary-mass object Cha1107-7626. We use spectroscopic data from VLT/XSHOOTER and from instruments on-board JWST to follow the evolution of the burst. Starting in late June 2025, the object exhibits enhanced line emission compared to previous epochs, including a much stronger H$\alpha$ feature. The H$\alpha$ line is double-peaked with red-shifted absorption, a hallmark for channeled, magnetospheric accretion. From the changes in Hydrogen lines, we infer an increase in the mass accretion rate by a factor of 6-8 (corresponding to $\sim$0.8-0.9 dex).

The optical continuum is elevated by a factor of 3-6 compared to quiescence. The near-infrared remains largely unchanged except for a $\sim$10\% increase in the final epoch, and the mid-infrared fluxes are enhanced by 10–20\%. While the silicate feature is effectively unchanged, we report clear changes in the hydrocarbon line spectrum during the burst, in particular the appearance of a feature at 6.6 $\mu$m attributed to water vapour. By the end of August 2025, the accretion burst is still ongoing and in fact shows the strongest accretion rates measured, mirrored by the strongest H$\alpha$ and optical continuum excess. This implies that the burst has lasted for more than 2 months, though its true duration remains unconstrained. The archival spectra for Cha1107-7626 indicate that a similar event may have happened about ten years ago.

Accretion variability is commonly observed in young stars, and has also been documented in substellar objects, including those with planetary masses \citep{viswanath2024}. For typical young stars, however, accretion rate changes do not exceed a factor of $\sim$3 on timescales of days to months \citep{nguyen2009}. Cha1107-7626 clearly shows unusually strong accretion variability compared to these typical cases. The duration of the observed event also distinguishes it from the most common variability of T Tauri stars, whose timescale is controlled by the stellar rotation \citep{costigan2012,costigan2014}. While the rotation period of Cha1107-7626 is not known, it can be expected to be in the range of 1-4\,days \citep{scholz2018}, which means that the accretion event lasted more than 15 rotational cycles. 

On the other hand, the observed event seen in Cha1107-7626 is comparable to EXor-type bursts, considering its duration, the optical amplitude, accretion rate increase, and the changes in the line spectrum \citep{fischer2023,sicilia2012}. As an example, the 2022 burst of the prototype EX Lupi lasted about four months. At its peak, the star was brightened by 2\,mag in the optical, with an increase in mass accretion by a factor of 7 \citep{cruz2023,singh2024}. The event observed for Cha1107-7626 shows very similar characteristics. Cha1107-7626 is therefore the first planetary-mass object identified as an EXor, extending this class of bursts to the regime of giant planets.

This kind of accretion bursts are key events in the early evolution of stars \citep{audard2014,fischer2023}. In particular, such events can have a significant effect on chemical and physical evolution of the disk \citep{abraham2009,smith2025}, and potentially on the early stages of planet formation. Our target is the lowest mass object observed thus far that is going through an accretion burst, and by far the lowest in the EXor category. Detailed studies of accretion variability have in the past helped to illuminate the interactions between young stellar objects and their disks, including the role of magnetic fields. Similarly, the observations presented here provide a glimpse into the nature of accretion in planetary-mass objects. 


\begin{acknowledgments}

\end{acknowledgments}

We thank John Pritchard and Paula Sanchez Saez at ESO for their support during the preparation and execution of the XSHOOTER observations. We also thank Weston Eck, Tony Keyes, and Greg Sloan at the Space Telescope Science Institute for their assistance with the JWST observations. VAA acknowledges support from the INAF grants 1.05.12.05.03 and 1.05.24.07.02. AS and BD acknowledge support from the UKRI Science and Technology Facilities Council through grant ST/Y001419/1/. AB acknowledges support from the Deutsche Forschungsgemeinschaft (DFG, German Research Foundation) under Germany's Excellence Strategy – EXC 2094 – 390783311. KM acknowledges support from the Fundação para a Ciência e a Tecnologia (FCT) through the grant 2022.03809.CEECIND. PP acknowledges funding from the UK Research and Innovation (UKRI) under the UK government’s Horizon Europe funding guarantee from ERC (under grant agreement No 101076489).

Based on observations collected at the European Southern Observatory under programmes 115.2850, 115.29FC and 115.29G3. This work is based [in part] on observations made with the NASA/ESA/CSA James Webb Space Telescope. The data were obtained from the Mikulski Archive for Space Telescopes at the Space Telescope Science Institute, which is operated by the Association of Universities for Research in Astronomy, Inc., under NASA contract NAS 5-03127 for JWST. These observations are associated with programs \#4583 and \#9488. The specific observations analyzed can be accessed via \dataset[10.17909/jfng-v154]{http://dx.doi.org/10.17909/jfng-v154}.

\facilities{JWST, ESO-VLT}

\software{EsoReflex \citep{freudling2013}, Molecfit \citep{smette2015}, Astropy \citep{astropy2013,astropy2018,astropy2022}, Matplotlib \citep{hunter2007}, NumPy \citep{harris2020}, SciPy \citep{virtanen2020}}

\bibliography{chaJ1107_burst}{}

\begin{thebibliography}{}
\expandafter\ifx\csname natexlab\endcsname\relax\def\natexlab#1{#1}\fi
\providecommand{\url}[1]{\href{#1}{#1}}
\providecommand{\dodoi}[1]{doi:~\href{http://doi.org/#1}{\nolinkurl{#1}}}
\providecommand{\doeprint}[1]{\href{http://ascl.net/#1}{\nolinkurl{http://ascl.net/#1}}}
\providecommand{\doarXiv}[1]{\href{https://arxiv.org/abs/#1}{\nolinkurl{https://arxiv.org/abs/#1}}}

\bibitem[{P. {{\'A}brah{\'a}m} {et~al.}(2009){{\'A}brah{\'a}m}, {Juh{\'a}sz}, {Dullemond}, {K{\'o}sp{\'a}l}, {van Boekel}, {Bouwman}, {Henning}, {Mo{\'o}r}, {Mosoni}, {Sicilia-Aguilar}, \& {Sipos}}]{abraham2009}
{{\'A}brah{\'a}m}, P., {Juh{\'a}sz}, A., {Dullemond}, C.~P., {et~al.} 2009, \bibinfo{title}{{Episodic formation of cometary material in the outburst of a young Sun-like star},} \nat, 459, 224, \dodoi{10.1038/nature08004}

\bibitem[{J.~M. {Alcal{\'a}} {et~al.}(2017){Alcal{\'a}}, {Manara}, {Natta}, {Frasca}, {Testi}, {Nisini}, {Stelzer}, {Williams}, {Antoniucci}, {Biazzo}, {Covino}, {Esposito}, {Getman}, \& {Rigliaco}}]{alcala2017}
{Alcal{\'a}}, J.~M., {Manara}, C.~F., {Natta}, A., {et~al.} 2017, \bibinfo{title}{{X-shooter spectroscopy of young stellar objects in Lupus. Accretion properties of class II and transitional objects},} \aap, 600, A20, \dodoi{10.1051/0004-6361/201629929}

\bibitem[{V. {Almendros-Abad} {et~al.}(2022){Almendros-Abad}, {Mu{\v{z}}i{\'c}}, {Moitinho}, {Krone-Martins}, \& {Kubiak}}]{almendros2022}
{Almendros-Abad}, V., {Mu{\v{z}}i{\'c}}, K., {Moitinho}, A., {Krone-Martins}, A., \& {Kubiak}, K. 2022, \bibinfo{title}{{Youth analysis of near-infrared spectra of young low-mass stars and brown dwarfs},} \aap, 657, A129, \dodoi{10.1051/0004-6361/202142050}

\bibitem[{Y. {Aoyama} {et~al.}(2021){Aoyama}, {Marleau}, {Ikoma}, \& {Mordasini}}]{aoyama2021}
{Aoyama}, Y., {Marleau}, G.-D., {Ikoma}, M., \& {Mordasini}, C. 2021, \bibinfo{title}{{Comparison of Planetary H{\ensuremath{\alpha}}-emission Models: A New Correlation with Accretion Luminosity},} \apjl, 917, L30, \dodoi{10.3847/2041-8213/ac19bd}

\bibitem[{A.~M. {Arabhavi} {et~al.}(2025){Arabhavi}, {Kamp}, {van Dishoeck}, {Henning}, {Jang}, {Christiaens}, {Gasman}, {Pascucci}, {Perotti}, {Grant}, {Barrado}, {G{\"u}del}, {Lagage}, {Caratti o Garatti}, {Lahuis}, {Waters}, {Kaeufer}, {Kanwar}, {Morales-Calder{\'o}n}, {Schwarz}, {Sellek}, {Tabone}, {Temmink}, \& {Vlasblom}}]{arabhavi2025}
{Arabhavi}, A.~M., {Kamp}, I., {van Dishoeck}, E.~F., {et~al.} 2025, \bibinfo{title}{{MINDS: The Very Low-mass Star and Brown Dwarf Sample Hidden Water in Carbon-dominated Protoplanetary Disks},} \apjl, 984, L62, \dodoi{10.3847/2041-8213/adc692}

\bibitem[{C. {Aspin} {et~al.}(2010){Aspin}, {Reipurth}, {Herczeg}, \& {Capak}}]{aspin2010}
{Aspin}, C., {Reipurth}, B., {Herczeg}, G.~J., \& {Capak}, P. 2010, \bibinfo{title}{{The 2008 Extreme Outburst of the Young Eruptive Variable Star EX Lupi},} \apjl, 719, L50, \dodoi{10.1088/2041-8205/719/1/L50}

\bibitem[{ {Astropy Collaboration} {et~al.}(2013){Astropy Collaboration}, {Robitaille}, {Tollerud}, {Greenfield}, {Droettboom}, {Bray}, {Aldcroft}, {Davis}, {Ginsburg}, {Price-Whelan}, {Kerzendorf}, {Conley}, {Crighton}, {Barbary}, {Muna}, {Ferguson}, {Grollier}, {Parikh}, {Nair}, {Unther}, {Deil}, {Woillez}, {Conseil}, {Kramer}, {Turner}, {Singer}, {Fox}, {Weaver}, {Zabalza}, {Edwards}, {Azalee Bostroem}, {Burke}, {Casey}, {Crawford}, {Dencheva}, {Ely}, {Jenness}, {Labrie}, {Lim}, {Pierfederici}, {Pontzen}, {Ptak}, {Refsdal}, {Servillat}, \& {Streicher}}]{astropy2013}
{Astropy Collaboration}, {Robitaille}, T.~P., {Tollerud}, E.~J., {et~al.} 2013, \bibinfo{title}{{Astropy: A community Python package for astronomy},} \aap, 558, A33, \dodoi{10.1051/0004-6361/201322068}

\bibitem[{ {Astropy Collaboration} {et~al.}(2018){Astropy Collaboration}, {Price-Whelan}, {Sip{\H{o}}cz}, {G{\"u}nther}, {Lim}, {Crawford}, {Conseil}, {Shupe}, {Craig}, {Dencheva}, {Ginsburg}, {VanderPlas}, {Bradley}, {P{\'e}rez-Su{\'a}rez}, {de Val-Borro}, {Aldcroft}, {Cruz}, {Robitaille}, {Tollerud}, {Ardelean}, {Babej}, {Bach}, {Bachetti}, {Bakanov}, {Bamford}, {Barentsen}, {Barmby}, {Baumbach}, {Berry}, {Biscani}, {Boquien}, {Bostroem}, {Bouma}, {Brammer}, {Bray}, {Breytenbach}, {Buddelmeijer}, {Burke}, {Calderone}, {Cano Rodr{\'\i}guez}, {Cara}, {Cardoso}, {Cheedella}, {Copin}, {Corrales}, {Crichton}, {D'Avella}, {Deil}, {Depagne}, {Dietrich}, {Donath}, {Droettboom}, {Earl}, {Erben}, {Fabbro}, {Ferreira}, {Finethy}, {Fox}, {Garrison}, {Gibbons}, {Goldstein}, {Gommers}, {Greco}, {Greenfield}, {Groener}, {Grollier}, {Hagen}, {Hirst}, {Homeier}, {Horton}, {Hosseinzadeh}, {Hu}, {Hunkeler}, {Ivezi{\'c}}, {Jain}, {Jenness}, {Kanarek}, {Kendrew}, {Kern}, {Kerzendorf}, {Khvalko}, {King}, {Kirkby}, {Kulkarni},
  {Kumar}, {Lee}, {Lenz}, {Littlefair}, {Ma}, {Macleod}, {Mastropietro}, {McCully}, {Montagnac}, {Morris}, {Mueller}, {Mumford}, {Muna}, {Murphy}, {Nelson}, {Nguyen}, {Ninan}, {N{\"o}the}, {Ogaz}, {Oh}, {Parejko}, {Parley}, {Pascual}, {Patil}, {Patil}, {Plunkett}, {Prochaska}, {Rastogi}, {Reddy Janga}, {Sabater}, {Sakurikar}, {Seifert}, {Sherbert}, {Sherwood-Taylor}, {Shih}, {Sick}, {Silbiger}, {Singanamalla}, {Singer}, {Sladen}, {Sooley}, {Sornarajah}, {Streicher}, {Teuben}, {Thomas}, {Tremblay}, {Turner}, {Terr{\'o}n}, {van Kerkwijk}, {de la Vega}, {Watkins}, {Weaver}, {Whitmore}, {Woillez}, {Zabalza}, \& {Astropy Contributors}}]{astropy2018}
{Astropy Collaboration}, {Price-Whelan}, A.~M., {Sip{\H{o}}cz}, B.~M., {et~al.} 2018, \bibinfo{title}{{The Astropy Project: Building an Open-science Project and Status of the v2.0 Core Package},} \aj, 156, 123, \dodoi{10.3847/1538-3881/aabc4f}

\bibitem[{ {Astropy Collaboration} {et~al.}(2022){Astropy Collaboration}, {Price-Whelan}, {Lim}, {Earl}, {Starkman}, {Bradley}, {Shupe}, {Patil}, {Corrales}, {Brasseur}, {N{\"o}the}, {Donath}, {Tollerud}, {Morris}, {Ginsburg}, {Vaher}, {Weaver}, {Tocknell}, {Jamieson}, {van Kerkwijk}, {Robitaille}, {Merry}, {Bachetti}, {G{\"u}nther}, {Aldcroft}, {Alvarado-Montes}, {Archibald}, {B{\'o}di}, {Bapat}, {Barentsen}, {Baz{\'a}n}, {Biswas}, {Boquien}, {Burke}, {Cara}, {Cara}, {Conroy}, {Conseil}, {Craig}, {Cross}, {Cruz}, {D'Eugenio}, {Dencheva}, {Devillepoix}, {Dietrich}, {Eigenbrot}, {Erben}, {Ferreira}, {Foreman-Mackey}, {Fox}, {Freij}, {Garg}, {Geda}, {Glattly}, {Gondhalekar}, {Gordon}, {Grant}, {Greenfield}, {Groener}, {Guest}, {Gurovich}, {Handberg}, {Hart}, {Hatfield-Dodds}, {Homeier}, {Hosseinzadeh}, {Jenness}, {Jones}, {Joseph}, {Kalmbach}, {Karamehmetoglu}, {Ka{\l}uszy{\'n}ski}, {Kelley}, {Kern}, {Kerzendorf}, {Koch}, {Kulumani}, {Lee}, {Ly}, {Ma}, {MacBride}, {Maljaars}, {Muna}, {Murphy}, {Norman},
  {O'Steen}, {Oman}, {Pacifici}, {Pascual}, {Pascual-Granado}, {Patil}, {Perren}, {Pickering}, {Rastogi}, {Roulston}, {Ryan}, {Rykoff}, {Sabater}, {Sakurikar}, {Salgado}, {Sanghi}, {Saunders}, {Savchenko}, {Schwardt}, {Seifert-Eckert}, {Shih}, {Jain}, {Shukla}, {Sick}, {Simpson}, {Singanamalla}, {Singer}, {Singhal}, {Sinha}, {Sip{\H{o}}cz}, {Spitler}, {Stansby}, {Streicher}, {{\v{S}}umak}, {Swinbank}, {Taranu}, {Tewary}, {Tremblay}, {de Val-Borro}, {Van Kooten}, {Vasovi{\'c}}, {Verma}, {de Miranda Cardoso}, {Williams}, {Wilson}, {Winkel}, {Wood-Vasey}, {Xue}, {Yoachim}, {Zhang}, {Zonca}, \& {Astropy Project Contributors}}]{astropy2022}
{Astropy Collaboration}, {Price-Whelan}, A.~M., {Lim}, P.~L., {et~al.} 2022, \bibinfo{title}{{The Astropy Project: Sustaining and Growing a Community-oriented Open-source Project and the Latest Major Release (v5.0) of the Core Package},} \apj, 935, 167, \dodoi{10.3847/1538-4357/ac7c74}

\bibitem[{M. {Audard} {et~al.}(2014){Audard}, {{\'A}brah{\'a}m}, {Dunham}, {Green}, {Grosso}, {Hamaguchi}, {Kastner}, {K{\'o}sp{\'a}l}, {Lodato}, {Romanova}, {Skinner}, {Vorobyov}, \& {Zhu}}]{audard2014}
{Audard}, M., {{\'A}brah{\'a}m}, P., {Dunham}, M.~M., {et~al.} 2014, \bibinfo{title}{{Episodic Accretion in Young Stars},} in Protostars and Planets VI, ed. H.~{Beuther}, R.~S. {Klessen}, C.~P. {Dullemond}, \& T.~{Henning}, 387--410, \dodoi{10.2458/azu_uapress_9780816531240-ch017}

\bibitem[{A. {Banzatti} {et~al.}(2025){Banzatti}, {Salyk}, {Pontoppidan}, {Carr}, {Zhang}, {Arulanantham}, {Krijt}, {{\"O}berg}, {Cleeves}, {Najita}, {Pascucci}, {Blake}, {Romero-Mirza}, {Bergin}, {Cieza}, {Pinilla}, {Long}, {Mallaney}, {Xie}, {Waggoner}, {Kaeufer}, \& {The Jdiscs Collaboration}}]{banzatti2025}
{Banzatti}, A., {Salyk}, C., {Pontoppidan}, K.~M., {et~al.} 2025, \bibinfo{title}{{Water in Protoplanetary Disks with JWST-MIRI: Spectral Excitation Atlas and Radial Distribution from Temperature Diagnostic Diagrams and Doppler Mapping},} \aj, 169, 165, \dodoi{10.3847/1538-3881/ada962}

\bibitem[{S.~K. {Betti} {et~al.}(2023){Betti}, {Follette}, {Ward-Duong}, {Peck}, {Aoyama}, {Bary}, {Dacus}, {Edwards}, {Marleau}, {Mohamed}, {Palmo}, {Plunkett}, {Robinson}, \& {Wang}}]{betti2023}
{Betti}, S.~K., {Follette}, K.~B., {Ward-Duong}, K., {et~al.} 2023, \bibinfo{title}{{The Comprehensive Archive of Substellar and Planetary Accretion Rates},} \aj, 166, 262, \dodoi{10.3847/1538-3881/ad06b8}

\bibitem[{M. {Bonnefoy} {et~al.}(2014){Bonnefoy}, {Chauvin}, {Lagrange}, {Rojo}, {Allard}, {Pinte}, {Dumas}, \& {Homeier}}]{bonnefoy2014}
{Bonnefoy}, M., {Chauvin}, G., {Lagrange}, A.~M., {et~al.} 2014, \bibinfo{title}{{A library of near-infrared integral field spectra of young M-L dwarfs},} \aap, 562, A127, \dodoi{10.1051/0004-6361/201118270}

\bibitem[{J. {Bouvier} {et~al.}(2003){Bouvier}, {Grankin}, {Alencar}, {Dougados}, {Fern{\'a}ndez}, {Basri}, {Batalha}, {Guenther}, {Ibrahimov}, {Magakian}, {Melnikov}, {Petrov}, {Rud}, \& {Zapatero Osorio}}]{bouvier2003}
{Bouvier}, J., {Grankin}, K.~N., {Alencar}, S.~H.~P., {et~al.} 2003, \bibinfo{title}{{Eclipses by circumstellar material in the T Tauri star AA Tau. II. Evidence for non-stationary magnetospheric accretion},} \aap, 409, 169, \dodoi{10.1051/0004-6361:20030938}

\bibitem[{N. {Calvet} \& E. {Gullbring}(1998){Calvet} \& {Gullbring}}]{calvet1998}
{Calvet}, N., \& {Gullbring}, E. 1998, \bibinfo{title}{{The Structure and Emission of the Accretion Shock in T Tauri Stars},} \apj, 509, 802, \dodoi{10.1086/306527}

\bibitem[{L.~M. {Close} {et~al.}(2025){Close}, {van Capelleveen}, {Weible}, {Wagner}, {Haffert}, {Males}, {Ilyin}, {Kenworthy}, {Li}, {Long}, {Ertel}, {Ginski}, {Weinberger}, {Follette}, {Liberman}, {Twitchell}, {Johnson}, {Kueny}, {Apai}, {Doyon}, {Foster}, {Gasho}, {Van Gorkom}, {Guyon}, {Kautz}, {McLeod}, {McEwen}, {Pearce}, {Schatz}, {Hedglen}, {Wu}, {Isbell}, {Power}, {Carlson}, {Close}, {Tonucci}, \& {Mars}}]{close2025}
{Close}, L.~M., {van Capelleveen}, R.~F., {Weible}, G., {et~al.} 2025, \bibinfo{title}{{Wide Separation Planets In Time (WISPIT): Discovery of a Gap H$α$ Protoplanet WISPIT 2b with MagAO-X},} arXiv e-prints, arXiv:2508.19046.
\newblock \doarXiv{2508.19046}

\bibitem[{G. {Costigan} {et~al.}(2012){Costigan}, {Scholz}, {Stelzer}, {Ray}, {Vink}, \& {Mohanty}}]{costigan2012}
{Costigan}, G., {Scholz}, A., {Stelzer}, B., {et~al.} 2012, \bibinfo{title}{{LAMP: the long-term accretion monitoring programme of T Tauri stars in Chamaeleon I},} \mnras, 427, 1344, \dodoi{10.1111/j.1365-2966.2012.22008.x}

\bibitem[{G. {Costigan} {et~al.}(2014){Costigan}, {Vink}, {Scholz}, {Ray}, \& {Testi}}]{costigan2014}
{Costigan}, G., {Vink}, J.~S., {Scholz}, A., {Ray}, T., \& {Testi}, L. 2014, \bibinfo{title}{{Temperaments of young stars: rapid mass accretion rate changes in T Tauri and Herbig Ae stars},} \mnras, 440, 3444, \dodoi{10.1093/mnras/stu529}

\bibitem[{F. {Cruz-S{\'a}enz de Miera} {et~al.}(2023){Cruz-S{\'a}enz de Miera}, {K{\'o}sp{\'a}l}, {Abrah{\'a}m}, {Claes}, {Manara}, {Wendeborn}, {Fiorellino}, {Giannini}, {Nisini}, {Sicilia-Aguilar}, {Campbell-White}, {Alcal{\'a}}, {Banzatti}, {Szab{\'o}}, {Lykou}, {Antoniucci}, {Varga}, {Siwak}, {Park}, {Nagy}, \& {Kun}}]{cruz2023}
{Cruz-S{\'a}enz de Miera}, F., {K{\'o}sp{\'a}l}, {\'A}., {Abrah{\'a}m}, P., {et~al.} 2023, \bibinfo{title}{{Brightness and mass accretion rate evolution during the 2022 burst of EX Lupi},} \aap, 678, A88, \dodoi{10.1051/0004-6361/202347063}

\bibitem[{M.~C. {Cushing} {et~al.}(2006){Cushing}, {Roellig}, {Marley}, {Saumon}, {Leggett}, {Kirkpatrick}, {Wilson}, {Sloan}, {Mainzer}, {Van Cleve}, \& {Houck}}]{cushing2006}
{Cushing}, M.~C., {Roellig}, T.~L., {Marley}, M.~S., {et~al.} 2006, \bibinfo{title}{{A Spitzer Infrared Spectrograph Spectral Sequence of M, L, and T Dwarfs},} \apj, 648, 614, \dodoi{10.1086/505637}

\bibitem[{B. {Damian} {et~al.}(2025){Damian}, {Scholz}, {Jayawardhana}, {Almendros-Abad}, {Flagg}, {Mu{\v{z}}i{\'c}}, {Natta}, {Pinilla}, \& {Testi}}]{damian2025}
{Damian}, B., {Scholz}, A., {Jayawardhana}, R., {et~al.} 2025, \bibinfo{title}{{Spectroscopy of Free-floating Planetary-mass Objects and Their Disks with JWST},} \aj, 170, 127, \dodoi{10.3847/1538-3881/adea50}

\bibitem[{W.~J. {Fischer} {et~al.}(2023){Fischer}, {Hillenbrand}, {Herczeg}, {Johnstone}, {Kospal}, \& {Dunham}}]{fischer2023}
{Fischer}, W.~J., {Hillenbrand}, L.~A., {Herczeg}, G.~J., {et~al.} 2023, \bibinfo{title}{{Accretion Variability as a Guide to Stellar Mass Assembly},} in Astronomical Society of the Pacific Conference Series, Vol. 534, Protostars and Planets VII, ed. S.~{Inutsuka}, Y.~{Aikawa}, T.~{Muto}, K.~{Tomida}, \& M.~{Tamura}, 355, \dodoi{10.48550/arXiv.2203.11257}

\bibitem[{L. {Flagg} {et~al.}(2025){Flagg}, {Scholz}, {Almendros-Abad}, {Jayawardhana}, {Damian}, {Mu{\v{z}}i{\'c}}, {Natta}, {Pinilla}, \& {Testi}}]{flagg2025}
{Flagg}, L., {Scholz}, A., {Almendros-Abad}, V., {et~al.} 2025, \bibinfo{title}{{Detection of Hydrocarbons in the Disk around an Actively Accreting Planetary-mass Object},} \apj, 986, 200, \dodoi{10.3847/1538-4357/add71d}

\bibitem[{W. {Freudling} {et~al.}(2013){Freudling}, {Romaniello}, {Bramich}, {Ballester}, {Forchi}, {Garc{\'\i}a-Dabl{\'o}}, {Moehler}, \& {Neeser}}]{freudling2013}
{Freudling}, W., {Romaniello}, M., {Bramich}, D.~M., {et~al.} 2013, \bibinfo{title}{{Automated data reduction workflows for astronomy. The ESO Reflex environment},} \aap, 559, A96, \dodoi{10.1051/0004-6361/201322494}

\bibitem[{C.~R. Harris {et~al.}(2020)Harris, Millman, van~der Walt, Gommers, Virtanen, Cournapeau, Wieser, Taylor, Berg, Smith, Kern, Picus, Hoyer, van Kerkwijk, Brett, Haldane, del R{\'{i}}o, Wiebe, Peterson, G{\'{e}}rard-Marchant, Sheppard, Reddy, Weckesser, Abbasi, Gohlke, \& Oliphant}]{harris2020}
Harris, C.~R., Millman, K.~J., van~der Walt, S.~J., {et~al.} 2020, \bibinfo{title}{Array programming with {NumPy},} Nature, 585, 357, \dodoi{10.1038/s41586-020-2649-2}

\bibitem[{T. {Henning} {et~al.}(2024){Henning}, {Kamp}, {Samland}, {Arabhavi}, {Kanwar}, {van Dishoeck}, {G{\"u}del}, {Lagage}, {Waelkens}, {Abergel}, {Absil}, {Barrado}, {Boccaletti}, {Bouwman}, {Caratti o Garatti}, {Geers}, {Glauser}, {Lahuis}, {Mueller}, {Nehm{\'e}}, {Olofsson}, {Pantin}, {Ray}, {Scheithauer}, {Vandenbussche}, {Waters}, {Wright}, {Argyriou}, {Christiaens}, {Franceschi}, {Gasman}, {Grant}, {Guadarrama}, {Jang}, {Morales-Calder{\'o}n}, {Pawellek}, {Perotti}, {Rodgers-Lee}, {Schreiber}, {Schwarz}, {Tabone}, {Temmink}, {Vlasblom}, {Colina}, {Greve}, \& {{\"O}stlin}}]{henning2024}
{Henning}, T., {Kamp}, I., {Samland}, M., {et~al.} 2024, \bibinfo{title}{{MINDS: The JWST MIRI Mid-INfrared Disk Survey},} \pasp, 136, 054302, \dodoi{10.1088/1538-3873/ad3455}

\bibitem[{J.~D. Hunter(2007)Hunter}]{hunter2007}
Hunter, J.~D. 2007, \bibinfo{title}{Matplotlib: A 2D Graphics Environment,} Computing in Science \& Engineering, 9, 90, \dodoi{10.1109/MCSE.2007.55}

\bibitem[{P. {Jakobsen} {et~al.}(2022){Jakobsen}, {Ferruit}, {Alves de Oliveira}, {Arribas}, {Bagnasco}, {Barho}, {Beck}, {Birkmann}, {B{\"o}ker}, {Bunker}, {Charlot}, {de Jong}, {de Marchi}, {Ehrenwinkler}, {Falcolini}, {Fels}, {Franx}, {Franz}, {Funke}, {Giardino}, {Gnata}, {Holota}, {Honnen}, {Jensen}, {Jentsch}, {Johnson}, {Jollet}, {Karl}, {Kling}, {K{\"o}hler}, {Kolm}, {Kumari}, {Lander}, {Lemke}, {L{\'o}pez-Caniego}, {L{\"u}tzgendorf}, {Maiolino}, {Manjavacas}, {Marston}, {Maschmann}, {Maurer}, {Messerschmidt}, {Moseley}, {Mosner}, {Mott}, {Muzerolle}, {Pirzkal}, {Pittet}, {Plitzke}, {Posselt}, {Rapp}, {Rauscher}, {Rawle}, {Rix}, {R{\"o}del}, {Rumler}, {Sabbi}, {Salvignol}, {Schmid}, {Sirianni}, {Smith}, {Strada}, {te Plate}, {Valenti}, {Wettemann}, {Wiehe}, {Wiesmayer}, {Willott}, {Wright}, {Zeidler}, \& {Zincke}}]{jakobson2022}
{Jakobsen}, P., {Ferruit}, P., {Alves de Oliveira}, C., {et~al.} 2022, \bibinfo{title}{{The Near-Infrared Spectrograph (NIRSpec) on the James Webb Space Telescope. I. Overview of the instrument and its capabilities},} \aap, 661, A80, \dodoi{10.1051/0004-6361/202142663}

\bibitem[{R. {Jayawardhana} \& V.~D. {Ivanov}(2006){Jayawardhana} \& {Ivanov}}]{jaya2006}
{Jayawardhana}, R., \& {Ivanov}, V.~D. 2006, \bibinfo{title}{{Spectroscopy of Young Planetary Mass Candidates with Disks},} \apjl, 647, L167, \dodoi{10.1086/507522}

\bibitem[{V. {Joergens} {et~al.}(2013){Joergens}, {Bonnefoy}, {Liu}, {Bayo}, {Wolf}, {Chauvin}, \& {Rojo}}]{joergens2013}
{Joergens}, V., {Bonnefoy}, M., {Liu}, Y., {et~al.} 2013, \bibinfo{title}{{OTS 44: Disk and accretion at the planetary border},} \aap, 558, L7, \dodoi{10.1051/0004-6361/201322432}

\bibitem[{A. {Juh{\'a}sz} {et~al.}(2012){Juh{\'a}sz}, {Dullemond}, {van Boekel}, {Bouwman}, {{\'A}brah{\'a}m}, {Acosta-Pulido}, {Henning}, {K{\'o}sp{\'a}l}, {Sicilia-Aguilar}, {Jones}, {Mo{\'o}r}, {Mosoni}, {Reg{\'a}ly}, {Szokoly}, \& {Sipos}}]{juhasz2012}
{Juh{\'a}sz}, A., {Dullemond}, C.~P., {van Boekel}, R., {et~al.} 2012, \bibinfo{title}{{The 2008 Outburst of EX Lup{\textemdash}Silicate Crystals in Motion},} \apj, 744, 118, \dodoi{10.1088/0004-637X/744/2/118}

\bibitem[{R. {Kurosawa} \& M.~M. {Romanova}(2013){Kurosawa} \& {Romanova}}]{kurosawa2013}
{Kurosawa}, R., \& {Romanova}, M.~M. 2013, \bibinfo{title}{{Spectral variability of classical T Tauri stars accreting in an unstable regime},} \mnras, 431, 2673, \dodoi{10.1093/mnras/stt365}

\bibitem[{A.~B. {Langeveld} {et~al.}(2024){Langeveld}, {Scholz}, {Mu{\v{z}}i{\'c}}, {Jayawardhana}, {Capela}, {Albert}, {Doyon}, {Flagg}, {de Furio}, {Johnstone}, {Lafr{\`e}niere}, \& {Meyer}}]{langeveld2024}
{Langeveld}, A.~B., {Scholz}, A., {Mu{\v{z}}i{\'c}}, K., {et~al.} 2024, \bibinfo{title}{{The JWST/NIRISS Deep Spectroscopic Survey for Young Brown Dwarfs and Free-floating Planets},} \aj, 168, 179, \dodoi{10.3847/1538-3881/ad6f0c}

\bibitem[{P.~W. {Lucas} \& P.~F. {Roche}(2000){Lucas} \& {Roche}}]{lucas2000}
{Lucas}, P.~W., \& {Roche}, P.~F. 2000, \bibinfo{title}{{A population of very young brown dwarfs and free-floating planets in Orion},} \mnras, 314, 858, \dodoi{10.1046/j.1365-8711.2000.03515.x}

\bibitem[{K.~L. {Luhman} {et~al.}(2005){Luhman}, {Adame}, {D'Alessio}, {Calvet}, {Hartmann}, {Megeath}, \& {Fazio}}]{luhman2005}
{Luhman}, K.~L., {Adame}, L., {D'Alessio}, P., {et~al.} 2005, \bibinfo{title}{{Discovery of a Planetary-Mass Brown Dwarf with a Circumstellar Disk},} \apjl, 635, L93, \dodoi{10.1086/498868}

\bibitem[{K.~L. {Luhman} {et~al.}(2008){Luhman}, {Allen}, {Allen}, {Gutermuth}, {Hartmann}, {Mamajek}, {Megeath}, {Myers}, \& {Fazio}}]{luhman2008}
{Luhman}, K.~L., {Allen}, L.~E., {Allen}, P.~R., {et~al.} 2008, \bibinfo{title}{{The Disk Population of the Chamaeleon I Star-forming Region},} \apj, 675, 1375, \dodoi{10.1086/527347}

\bibitem[{N. {Miret-Roig}(2023){Miret-Roig}}]{miret2023}
{Miret-Roig}, N. 2023, \bibinfo{title}{{The origin of free-floating planets},} \apss, 368, 17, \dodoi{10.1007/s10509-023-04175-5}

\bibitem[{A. {Modigliani} {et~al.}(2010){Modigliani}, {Goldoni}, {Royer}, {Haigron}, {Guglielmi}, {Fran{\c{c}}ois}, {Horrobin}, {Bristow}, {Vernet}, {Moehler}, {Kerber}, {Ballester}, {Mason}, \& {Christensen}}]{modigliani2010}
{Modigliani}, A., {Goldoni}, P., {Royer}, F., {et~al.} 2010, \bibinfo{title}{{The X-shooter pipeline},} in Society of Photo-Optical Instrumentation Engineers (SPIE) Conference Series, Vol. 7737, Observatory Operations: Strategies, Processes, and Systems III, ed. D.~R. {Silva}, A.~B. {Peck}, \& B.~T. {Soifer}, 773728, \dodoi{10.1117/12.857211}

\bibitem[{M. {Morales-Calder{\'o}n} {et~al.}(2025){Morales-Calder{\'o}n}, {Jang}, {Arabhavi}, {Christiaens}, {Barrado}, {Kamp}, {van Dishoeck}, {Henning}, {Waters}, {Temmink}, {G{\"u}del}, {Lagage}, {Garatti}, {Glauser}, {Ray}, {Franceschi}, {Gasman}, {Grant}, {Kaeufer}, {Kanwar}, {Perotti}, {Samland}, {Schwarz}, {Vlasblom}, {Colina}, \& {{\"O}stlin}}]{morales2025}
{Morales-Calder{\'o}n}, M., {Jang}, H., {Arabhavi}, A.~M., {et~al.} 2025, \bibinfo{title}{{MINDS. Cha H{\ensuremath{\alpha}} 1, a brown dwarf with a hydrocarbon-rich disk},} arXiv e-prints, arXiv:2508.05155, \dodoi{10.48550/arXiv.2508.05155}

\bibitem[{D.~C. {Nguyen} {et~al.}(2009){Nguyen}, {Scholz}, {van Kerkwijk}, {Jayawardhana}, \& {Brandeker}}]{nguyen2009}
{Nguyen}, D.~C., {Scholz}, A., {van Kerkwijk}, M.~H., {Jayawardhana}, R., \& {Brandeker}, A. 2009, \bibinfo{title}{{How Variable is Accretion in Young Stars?},} \apjl, 694, L153, \dodoi{10.1088/0004-637X/694/2/L153}

\bibitem[{B.~A. {Sargent} {et~al.}(2014){Sargent}, {Forrest}, {Watson}, {D'Alessio}, {Calvet}, {Furlan}, {Kim}, {Green}, {Pontoppidan}, {Richter}, \& {Tayrien}}]{sargent2014}
{Sargent}, B.~A., {Forrest}, W., {Watson}, D.~M., {et~al.} 2014, \bibinfo{title}{{Emission from Water Vapor and Absorption from Other Gases at 5-7.5 {\ensuremath{\mu}}m in Spitzer-IRS Spectra of Protoplanetary Disks},} \apj, 792, 83, \dodoi{10.1088/0004-637X/792/2/83}

\bibitem[{A. {Scholz} {et~al.}(2013){Scholz}, {Froebrich}, \& {Wood}}]{scholz2013}
{Scholz}, A., {Froebrich}, D., \& {Wood}, K. 2013, \bibinfo{title}{{A systematic survey for eruptive young stellar objects using mid-infrared photometry},} \mnras, 430, 2910, \dodoi{10.1093/mnras/stt091}

\bibitem[{A. {Scholz} \& R. {Jayawardhana}(2006){Scholz} \& {Jayawardhana}}]{scholz2006}
{Scholz}, A., \& {Jayawardhana}, R. 2006, \bibinfo{title}{{Variable Accretion and Outflow in Young Brown Dwarfs},} \apj, 638, 1056, \dodoi{10.1086/498973}

\bibitem[{A. {Scholz} {et~al.}(2005){Scholz}, {Jayawardhana}, \& {Brandeker}}]{scholz2005}
{Scholz}, A., {Jayawardhana}, R., \& {Brandeker}, A. 2005, \bibinfo{title}{{Whims of an Accreting Young Brown Dwarf: Exploring the Emission-Line Variability of 2MASSW J1207334-393254},} \apjl, 629, L41, \dodoi{10.1086/444358}

\bibitem[{A. {Scholz} {et~al.}(2018){Scholz}, {Moore}, {Jayawardhana}, {Aigrain}, {Peterson}, \& {Stelzer}}]{scholz2018}
{Scholz}, A., {Moore}, K., {Jayawardhana}, R., {et~al.} 2018, \bibinfo{title}{{A Universal Spin-Mass Relation for Brown Dwarfs and Planets},} \apj, 859, 153, \dodoi{10.3847/1538-4357/aabfbe}

\bibitem[{A. {Scholz} {et~al.}(2022){Scholz}, {Muzic}, {Jayawardhana}, {Quinlan}, \& {Wurster}}]{scholz2022}
{Scholz}, A., {Muzic}, K., {Jayawardhana}, R., {Quinlan}, L., \& {Wurster}, J. 2022, \bibinfo{title}{{Rogue Planets and Brown Dwarfs: Predicting the Populations Free-floating Planetary Mass Objects Observable with JWST},} \pasp, 134, 104401, \dodoi{10.1088/1538-3873/ac9431}

\bibitem[{H.~H. {Seo} \& A. {Scholz}(2025){Seo} \& {Scholz}}]{seo2025}
{Seo}, H.~H., \& {Scholz}, A. 2025, \bibinfo{title}{{Discs around young free-floating planetary-mass objects: ultradeep Spitzer imaging of IC348},} \mnras, 537, 2579, \dodoi{10.1093/mnras/staf163}

\bibitem[{A. {Sicilia-Aguilar} {et~al.}(2012){Sicilia-Aguilar}, {K{\'o}sp{\'a}l}, {Setiawan}, {{\'A}brah{\'a}m}, {Dullemond}, {Eiroa}, {Goto}, {Henning}, \& {Juh{\'a}sz}}]{sicilia2012}
{Sicilia-Aguilar}, A., {K{\'o}sp{\'a}l}, {\'A}., {Setiawan}, J., {et~al.} 2012, \bibinfo{title}{{Optical spectroscopy of EX Lupi during quiescence and outburst. Infall, wind, and dynamics in the accretion flow},} \aap, 544, A93, \dodoi{10.1051/0004-6361/201118555}

\bibitem[{K. {Singh} {et~al.}(2024){Singh}, {Ninan}, {Romanova}, {Buckley}, {Ojha}, {Ghosh}, {Monson}, {Schramm}, {Sharma}, {Reichart}, {Mikolajewska}, {Beamin}, {Borissova}, {Ivanov}, {Kouprianov}, {Hambsch}, \& {Pearce}}]{singh2024}
{Singh}, K., {Ninan}, J.~P., {Romanova}, M.~M., {et~al.} 2024, \bibinfo{title}{{Accretion Funnel Reconfiguration during an Outburst in a Young Stellar Object: EX Lupi},} \apj, 968, 88, \dodoi{10.3847/1538-4357/ad4099}

\bibitem[{A. {Smette} {et~al.}(2015){Smette}, {Sana}, {Noll}, {Horst}, {Kausch}, {Kimeswenger}, {Barden}, {Szyszka}, {Jones}, {Gallenne}, {Vinther}, {Ballester}, \& {Taylor}}]{smette2015}
{Smette}, A., {Sana}, H., {Noll}, S., {et~al.} 2015, \bibinfo{title}{{Molecfit: A general tool for telluric absorption correction. I. Method and application to ESO instruments},} \aap, 576, A77, \dodoi{10.1051/0004-6361/201423932}

\bibitem[{S.~A. {Smith} {et~al.}(2025){Smith}, {Romero-Mirza}, {Banzatti}, {Rab}, {{\'A}brah{\'a}m}, {K{\'o}sp{\'a}l}, {Claes}, {Manara}, {{\"O}berg}, {Bouwman}, {de Miera}, \& {Green}}]{smith2025}
{Smith}, S.~A., {Romero-Mirza}, C.~E., {Banzatti}, A., {et~al.} 2025, \bibinfo{title}{{JWST's Sharper View of EX Lup: Cold Water from Ice Sublimation during Accretion Outbursts},} \apjl, 984, L51, \dodoi{10.3847/2041-8213/adcab8}

\bibitem[{L. {Testi} {et~al.}(2002){Testi}, {Natta}, {Oliva}, {D'Antona}, {Comeron}, {Baffa}, {Comoretto}, \& {Gennari}}]{testi2002}
{Testi}, L., {Natta}, A., {Oliva}, E., {et~al.} 2002, \bibinfo{title}{{A Young Very Low Mass Object Surrounded by Warm Dust},} \apjl, 571, L155, \dodoi{10.1086/341361}

\bibitem[{P. Virtanen {et~al.}(2020)Virtanen, Gommers, Oliphant, Haberland, Reddy, Cournapeau, Burovski, Peterson, Weckesser, Bright, {van der Walt}, Brett, Wilson, Millman, Mayorov, Nelson, Jones, Kern, Larson, Carey, Polat, Feng, Moore, {VanderPlas}, Laxalde, Perktold, Cimrman, Henriksen, Quintero, Harris, Archibald, Ribeiro, Pedregosa, {van Mulbregt}, \& {SciPy 1.0 Contributors}}]{virtanen2020}
Virtanen, P., Gommers, R., Oliphant, T.~E., {et~al.} 2020, \bibinfo{title}{{{SciPy} 1.0: Fundamental Algorithms for Scientific Computing in Python},} Nature Methods, 17, 261, \dodoi{10.1038/s41592-019-0686-2}

\bibitem[{G. {Viswanath} {et~al.}(2024){Viswanath}, {Ringqvist}, {Demars}, {Janson}, {Bonnefoy}, {Aoyama}, {Marleau}, {Dougados}, {Szul{\'a}gyi}, \& {Thanathibodee}}]{viswanath2024}
{Viswanath}, G., {Ringqvist}, S.~C., {Demars}, D., {et~al.} 2024, \bibinfo{title}{{ExoplaNeT accRetion mOnitoring sPectroscopic surveY (ENTROPY): I. Evidence for magnetospheric accretion in the young isolated planetary-mass object 2MASS J11151597+1937266},} \aap, 691, A64, \dodoi{10.1051/0004-6361/202450881}

\bibitem[{G.~S. {Wright} {et~al.}(2023){Wright}, {Rieke}, {Glasse}, {Ressler}, {Garc{\'\i}a Mar{\'\i}n}, {Aguilar}, {Alberts}, {{\'A}lvarez-M{\'a}rquez}, {Argyriou}, {Banks}, {Baudoz}, {Boccaletti}, {Bouchet}, {Bouwman}, {Brandl}, {Breda}, {Bright}, {Cale}, {Colina}, {Cossou}, {Coulais}, {Cracraft}, {De Meester}, {Dicken}, {Engesser}, {Etxaluze}, {Fox}, {Friedman}, {Fu}, {Gasman}, {G{\'a}sp{\'a}r}, {Gastaud}, {Geers}, {Glauser}, {Gordon}, {Greene}, {Greve}, {Grundy}, {G{\"u}del}, {Guillard}, {Haderlein}, {Hashimoto}, {Henning}, {Hines}, {Holler}, {Detre}, {Jahromi}, {James}, {Jones}, {Justtanont}, {Kavanagh}, {Kendrew}, {Klaassen}, {Krause}, {Labiano}, {Lagage}, {Lambros}, {Larson}, {Law}, {Lee}, {Libralato}, {Lorenzo Alverez}, {Meixner}, {Morrison}, {Mueller}, {Murray}, {Mycroft}, {Myers}, {Nayak}, {Naylor}, {Nickson}, {Noriega-Crespo}, {{\"O}stlin}, {O'Sullivan}, {Ottens}, {Patapis}, {Penanen}, {Pietraszkiewicz}, {Ray}, {Regan}, {Roteliuk}, {Royer}, {Samara-Ratna}, {Samuelson}, {Sargent}, {Scheithauer},
  {Schneider}, {Schreiber}, {Shaughnessy}, {Sheehan}, {Shivaei}, {Sloan}, {Tamas}, {Teague}, {Temim}, {Tikkanen}, {Tustain}, {van Dishoeck}, {Vandenbussche}, {Weilert}, {Whitehouse}, \& {Wolff}}]{wright2023}
{Wright}, G.~S., {Rieke}, G.~H., {Glasse}, A., {et~al.} 2023, \bibinfo{title}{{The Mid-infrared Instrument for JWST and Its In-flight Performance},} \pasp, 135, 048003, \dodoi{10.1088/1538-3873/acbe66}

\bibitem[{C. {Xie} {et~al.}(2023){Xie}, {Pascucci}, {Long}, {Pontoppidan}, {Banzatti}, {Kalyaan}, {Salyk}, {Liu}, {Najita}, {Pinilla}, {Arulanantham}, {Herczeg}, {Carr}, {Bergin}, {Ballering}, {Krijt}, {Blake}, {Zhang}, {{\"O}berg}, {Green}, \& {Jdiscs Collaboration}}]{xie2023}
{Xie}, C., {Pascucci}, I., {Long}, F., {et~al.} 2023, \bibinfo{title}{{Water-rich Disks around Late M Stars Unveiled: Exploring the Remarkable Case of Sz 114},} \apjl, 959, L25, \dodoi{10.3847/2041-8213/ad0ed9}

\bibitem[{M.~R. {Zapatero Osorio} {et~al.}(2000){Zapatero Osorio}, {B{\'e}jar}, {Mart{\'\i}n}, {Rebolo}, {Barrado y Navascu{\'e}s}, {Bailer-Jones}, \& {Mundt}}]{zapatero2000}
{Zapatero Osorio}, M.~R., {B{\'e}jar}, V.~J.~S., {Mart{\'\i}n}, E.~L., {et~al.} 2000, \bibinfo{title}{{Discovery of Young, Isolated Planetary Mass Objects in the {\ensuremath{\sigma}} Orionis Star Cluster},} Science, 290, 103, \dodoi{10.1126/science.290.5489.103}

\end{thebibliography}
\bibliographystyle{aasjournalv7}




\end{document}